\documentclass{article}
\usepackage[utf8]{inputenc}
\usepackage[a4paper,
            bindingoffset=0.0in,
            left=1in,
            right=1in,
            top=1in,
            bottom=1in,
            footskip=.25in]{geometry}
\usepackage{color,soul}
\usepackage[
backend=biber,
style=numeric,
sorting=none
]{biblatex}
\usepackage{verbatim}
\usepackage{siunitx}
\usepackage{graphicx}
\usepackage{subcaption}
\usepackage{authblk}
\usepackage{adjustbox}
\title{Beyond S-curves: Recurrent Neural Networks for Technology Forecasting}

\author[1, 2]{Alexander Glavackij}
\author[2, 3, 4]{Dimitri Percia David}
\author[2]{Alain Mermoud}
\author[1]{Angelika Romanou}
\author[1]{Karl Aberer}
\affil[1]{EPFL, School of Computer and Communication Sciences}
\affil[2]{Cyber-Defence Campus, armasuisse Science and Technology}
\affil[3]{University of Geneva, Geneva School of Economics and Management}
\affil[4]{University of Applied Sciences of Western Switzerland, Institute of Entrepreneurship \& Management}

\begin{document}

\ifx
\section*{Feedback}

Comments from the Guest Editorial Board:

Thank you for considering our special issue. We have collected three reviewer reports, but unfortunately, we observed several critical comments that may lead us to decide not to move this submission forward.
\begin{enumerate}
    \item Despite a well-described motivation on technological forecasting with an RNN model on Arxiv data, our reviewers reminded us that this submission highlights three datasets. The main one using Arxiv data lacks sufficient interpretation and understanding on which specific technologies this work attempts to forecast. At the same time, the other two datasets on sales and hospital visitors are irrelevant to our interests.
    \item We appreciate the attempt to handle "technological forecasting" - a super interesting topic for us, but we noticed this work only slightly touched on this concept at a superficial level, without any solid theoretical support on building up the models (e.g., why choosing those parameters, and why do you consider the connections between those selected parameters and the to-be-forecast technologies are solid and convincing).
    \item Again, for "technological forecasting" studies, we super emphasize a good story to interpret your forecasting results and their broad implications. Even though this might not be a key focus of our special issue and the journal, we should have appreciated such a story to easily connect the models with our tech-mining researchers/readers.
\end{enumerate}

Ps, since we have sent your invitation to participate our IP\&MC2022 conference, considering the outcome of this submission, we still welcome the authors present this work as a poster in our AI+Informetrics track.

Damiano Spina (Managing Guest Editor) Reject

\subsection*{Reviewer 1}

\begin{itemize}
    \item{Research summary:} This paper tries to build a neural network model to predict the number of publications on Arxiv. Though the authors claimed it to be a technology forcasting task, however, I failed to recognize the contribution to this specific task. So I recommend a rejection of this submission.
    \item{Major Strengths:} no
    \item{Major Weaknesses:} research objective is not clear
    literature review is not sufficient
    major contribution is not enough for publication
    \item{Grammar and Readability:} ok
    \item{Specific comments:} However, I still recommend the authors might improve the work and get it submitted to somewhere else following my comments below:
    First, I recommend the authors restate the objective of this paper. As I mentioned above, your major work in this study is to predict the number of publications uploaded to Arxiv in the future. You might rewrite your Introduction, literature review and the like based on this contribution.
    Second, I do not think you did a good work on your literature review. Please systematically review related studies to your work, not just enumerate publications. Meanwhile, if you were working on technology forcasting, you would include more publications on this topic, however I did not find very important citations from your draft. In addtion, please read author guideline carefully before submission. The in-text citations should follow an APA style.
    Third, in the methodology part, the authors included a lot technical discussion. I would recommend the authors move them to the literature review section if they are necessary.
    Last, concerning the major contribution of this task, when I checked Figure 6, I have found that if I drew a mean value of publication history line across and could still make a relative good performance as the model did. So I cannot see the value of this work.
    \item{Concluding remarks:} I recommend a rejection.
\end{itemize}

\subsection*{Reviewer 2}
\begin{enumerate}
    \item This paper develops an approach to technology forecasting. For this, a recursive neural network (RNN) machine learning model that predicts future counts of e-prints published in subcategories of ArXiv is designed and tested. This paper explores an interesting topic that is fitted to the scope of Information Processing \& Management. However, some parts should be elaborated further to emphasize the originality and contributions of the paper, as follows.
    \item Recent literature has developed data-driven approaches to technology forecasting. there are several studies that anticipate technology development by predicting the number of citations that patents would receive (see e.g., https://doi.org/10.1016/j.techfore.2017.10.002, https://doi.org/10.1016/j.joi.2017.03.007). To demonstrate the originality and contribution of this study, the authors need to introduce previous approaches to technology forecasting based on patent analytics. Moreover, the differences between the previous data-driven approaches to technology forecasting and the proposed approach need to be provided.
    \item Patent information, which contains vast and reliable information on technology developments, has been widely used to develop approaches to technology forecasting. In this respect, the authors need to elaborate on the comparative benefits of using ArXiv data relative to patent information for developing an approach to technology forecasting.
    \item It is stated that "… we train our model to predict the number of these same uploaded e-prints in any subcategory of arXiv - a proxy we use as a technology taxonomy." in Page 1. Here, the authors need to elaborate more on the meaning of "technology taxonomy". Moreover, relationships between subcategories of arXiv data and technologies need to be explained more clearly.
    \item The authors stated general benefits of using machine learning models for time series forecasting in Page 1. Since the purpose of this paper is to develop an approach to technology forecasting, the authors need to state the benefits of using machine learning models in technology forecasting context.
    \item The authors performed experiments using two time-series datasets: (1) jewelry sales and (2) number of hospital visitors. However, since the purpose of this paper is not to develop a time-series forecasting approach but to develop a technology forecasting approach, the performance of the proposed approach in forecasting jewelry sales and the number of hospital visitors is not directly related with the validity of the approach. In this respect, I suggest the authors to replace these experiments with the experiments using other technology databases.
    \item Previous studies have shown that word embedding model with large dimensions result in better word representations. In this context, the authors need to provide rationales for reducing the size of word embedding vectors from 768 to 15.
    \item Scientific publication data contains jargons that are specific to their specialized fields. Thus, fine-tuning the pretrained word embedding model with ArXiv data would generate more fine-grained word embedding model that may enhance the performance of technology forecasting. In this respect, I suggest the authors (1) fine-tune the pretrained word embedding model with ArXiv data and (2) investigate the performance of the proposed approach based on the fine-tuned model.
\end{enumerate}

Minor comments:
\begin{enumerate}
    \item The authors repeated two very similar sentences in Page 2: "Semantic information refers to the meaning the texts convey, for example, the technologies and topics appearing in the texts" and "Semantic information refers to the meaning conveyed by textual data, for example, topics mentioned in the texts."
    \item There is a typo in Page 5. "Vec2word model" needs to be rewritten as "Word2vec model".
\end{enumerate}

\subsection*{Reviewer 3}
\begin{itemize}
    \item{Research summary:} Recently, many studies have been conducted through deep learning-based approaches in the field of technology management including technology prediction. This study proposed a machine learning-based approach method using RNN for technology prediction. In this study, the authors designed an experiment reflecting short- and long-term perspectives by composing arXiv documents related to science and technology into a dataset, verifying the proposed model's performance.
    \item{Major Strengths:}It suggested an RNN-based technology forecasting model using academic journals as a dataset. It conducted various experiments considering time series, deep learning models, features, and so on. It also showed a model performance.
    \item{Major Weaknesses:} It is necessary to present the purpose of this study and the target task of the proposed model in detail. Also, from the managerial implication point of view, it seems necessary to supplement the contributions of this study in comparison with previous studies.
    \item{Grammar and Readability:} Please check them below
- (in introduction) a mean absolute term error (MAPE) -\> percentage?
- (in 3.2) vec2word -\> word2vec?
    \item{Specific comments:} My specific comments concerning this manuscript are:
    \begin{enumerate}
        \item (Introduction, first paragraph) The subject of this study, 'technology prediction', has a very broad scope. For example, a technology prediction problem can include predicting the lifetime of technology and predicting a value for specific technology characteristics (impact, market value, etc.). In this study, it is necessary to define more specific research questions from which perspective the research target was set among these broad technology predictions. In particular, the machine learning model-based approach needs to define specific variables. In this process, it is also necessary to describe the relevance of whether the conceptual task of technology prediction can be interpreted as a proxy for a specific variable. It seems required to supplement the introduction by specifying a detailed description of the purpose and target task.
        \item (Introduction, second paragraph) This study set up academic papers on e-prints as a dataset in the second paragraph. It can be questionable whether the agnostic to any technology or industry can be shown by selecting only one technology field and applying it to the proposed methodology. Although only the e-prints field is targeted, it can include different technical areas below it. Through these considerations, can you show agnostic to any technology or industry as mentioned in the text? If it is true, it seems that the explanation for this should be supplemented in order to secure the basis for what the author said.
        \item (Introduction, second paragraph) The following sentence does not make sense to me: 'This shows that similar subcategories on arXiv do not exhibit distinct development patterns that the model can use to forecast.' Why can't we use a predictive model when similar subcategories shouldn't show distinct development patterns?
        \item (In the chapter 2.1) This chapter presents two approaches to Technology Forecasting: Semantic and structural information. It needs to be offered based on the relevance to the method proposed in this study. It seems that this study did not use structural information. And semantic information was understood as being added on as an auxiliary feature. Then, it is necessary to explain why structural information is not appropriate for this study, and what points are more emphasized. I think the previous research reviews from both perspectives are not very relevant to this study. Whether or not the approach of technical forecasting done based on predictive models needs to be reviewed. It is recommended to consider whether studies with high relevance to this study should be reviewed, including what tasks were set to perform technical prediction, what data variables were defined, and what models were used.
        \item (In the chapter 3.2) Does the proposed model target well reconstructing time-series $z_i$ with encoder and decoder? Is it correct that time-series $z_i$ means 'the number of e-print uploads'? The model input contains the time-series $z_i$ and the corresponding co-variates $x_i$, what is the difference between the two variables? How is the objective function defined in model training? How are hyperparameters such as learning rate, layer unit, and optimize set? Which text field (title, abstract, claim, etc.) was used for text embedding among co-variates? Finally, if the model learns 'the number of e-print uploads', is it advantageous to use RNN for this model? A typical application of Seq-to-seq is a translation model between different languages. In predicting the number of papers based on time series, it seems that using regression-based ARIMA etc. may be more accurate.
        \item (In chapter 4) For the task of technology management, application and implication are important from the management point of view. In other words, it seems to be supplemented by the explanation of what the model presented in this study predicted. It seems necessary to interpret from the perspective of technology management, such as which technology fields will develop in the future in the E-print field.
    \end{enumerate}
\end{itemize}

\section*{TODO}
\begin{itemize}
    \item merge GTM paper and IP\&M papers
    \item Finish baseline analysis
    \item Define research question/insights taken from the paper properly and precisely
    \item Redo literature review, structure differently
    \item clarify link between predicting e-print uploads and technological development (link between research and patents?)
    \item find literature that establishes link between some sort of (economic) value created and the number of articles in a category
    \item justify choice of the ML model/architecture: why did we pick that one?
\end{itemize}

\section*{Main Review Points}
\begin{itemize}
    \item reorient literature review to include model forecasting approaches as given by reviewer 2
    \item explain why we use scientific literature instead of patents
    \item explain the relationship between arxiv taxonomy and technologies better
    \item drop validation datasets
\end{itemize}

\noindent
\textbf{Key message of this work: using S-curves to predict technological development is as accurate as model-based forecasting approaches.}

\newpage

\fi

\maketitle

\section*{Abstract}
\noindent
Because of the considerable heterogeneity and complexity of the technological landscape, building accurate models to forecast is a challenging endeavor.
Due to their high prevalence in many complex systems, S-curves are a popular forecasting approach in previous work.
However, their forecasting performance has not been directly compared to other technology forecasting approaches.
Additionally, recent developments in time series forecasting that claim to improve forecasting accuracy are yet to be applied to technological development data.
This work addresses both research gaps by comparing the forecasting performance of S-curves to a baseline and by developing an \emph{autencoder} approach that employs recent advances in machine learning and time series forecasting.
S-curves forecasts largely exhibit a mean average percentage error (MAPE) comparable to a simple ARIMA baseline.
However, for a minority of emerging technologies, the MAPE increases by two magnitudes.
Our autoencoder approach improves the MAPE by 13.5\% on average over the second-best result.
It forecasts established technologies with the same accuracy as the other approaches.
However, it is especially strong at forecasting emerging technologies with a mean MAPE 18\% lower than the next best result.
Our results imply that a simple ARIMA model is preferable over the S-curve for technology forecasting.
Practitioners looking for more accurate forecasts should opt for the presented autoencoder approach.
 
\section{Introduction}
\noindent
In an environment where technological development largely determines economic growth and social change, predicting these developments is critical for strategic decisions in organizations~\cite{schwab_the_2018}.
Predicting technological development necessitates understanding and modeling it~\cite{roadmapping_daim_2021}.
Once a model for technological development exists, one can use that model to forecast to aid strategic decisions \cite{hung_technological_2014} \cite{lai_structured_2017} \cite{zhou_deep_2021}.
The pay-off of having a model to forecast with sufficient accuracy is significant; however, constructing such models is difficult due to the considerable heterogeneity and the complexity of the technological landscape~\cite{huang_exploring_2021}~\cite{zhang_forecasting_2019}~\cite{huang_assessment_2019}.

Successful innovations become gradually more adopted by the market and society.
This diffusion is a gradual process that can be modeled by S-curves~\cite{rogers2014diffusion}~\cite{ayres_technological_1969}, which have a long history and appear in many different domains, e.g., Biology \cite{kingsland_refractory_1982}.
S-curve segment development into three phases: introduction, growth, and maturation phases~\cite{priestley_innovation_2020} \cite{lotfi_forecasting_2014}.
They help explain past observed development behavior.

However, we find several pitfalls when using S-curves to forecast technological development.
First, forecasting is the most useful for emerging technologies, i.e., in the Introduction phase of the S-curve.
However, little data on the technology's development is available in this phase.
This poses problems for S-curve forecasting: accurate S-curve forecasts require more data, only available at a later life cycle stage where more of the technology's development has been observed.
However, at that point, the utility of forecasts is reduced, as most of the technology's development has already happened.
Second, to correctly forecast using S-curves, practitioners must assume that the development will follow an S-curve pattern.
%However, there is insufficient information available for emerging technologies to make this assumption.
To our knowledge, quantitative works have yet to empirically confirm that such an S-curve is eventually observed across all technology domains (especially not done in information technologies, which have a reduced or accelerated life cycle).

Recent developments in machine learning methods address these pitfalls.
They can capture complex non-linear relationships between input and output indicators and are thus especially strong at finding patterns in large amounts of data~\cite{vemuri_hundred-page_2020}. 
They claim to improve forecasting capabilities by reducing the assumptions taken by finding patterns in the data in a supervised or unsupervised manner, i.e., they are data-driven.

The first of our two-fold contribution is an evaluation of the forecasting capabilities of the S-curve.
We found a research gap in comparing the S-curve's forecasting capabilities against baselines.
To address this gap, we develop a practitioner's guide to S-curve forecasting and compare the forecasts' accuracy to an ARIMA baseline.
We find that S-curves largely exhibit a similar forecasting accuracy to the ARIMA approach except for a small minority of emerging technologies, where the forecasting error increases.

The second of our two-fold contribution is a forecasting approach leveraging recurrent neural networks (RNNs) to forecast technological development.
By using a modern data-driven approach, we aim to address the earlier described pitfalls of S-curves and to improve forecasting accuracy for technologies early in the life cycle.
We apply a sequence-to-sequence model developed by Salinas et al. that forecasts product demand on Amazon to technological forecasting~\cite{salinas_deepar_2019}.
This approach improves out-of-sample mean average percentage error (MAPE) by 14\% on average and by 12\% (median) for emerging technologies over the next best result.

Our data source for this work is arXiv, a platform that provides access to more than two million scholarly works in over 170 technical domains.
Most previous work uses patents as their data source, as a direct link between patents and economic value has been shown~\cite{gambardella_value_2008}~\cite{schmoch_indicators_1997}.
Previous work has established strong links between scholarly works and patents~\cite{david_is_2000}~\cite{cohen_links_2002}~\cite{rothaermel_university_2007}, and scholarly articles predate patents by four to five years~\cite{segev_analysis_2015}.
Thus, we use arXiv as our data source to forecast trends at an earlier stage than would be possible by using patents.

We structure the remainder of this article as follows.
In Section~\ref{sec2}, we justify using scholarly works to forecast technological development in previous work and then give an overview of related forecasting approaches.
Section~\ref{sec3} presents our dataset and how we produce comparable forecasts with the forecasting approaches we evaluate in this work.
In Section~\ref{sec4}, we present our results, give an interpretation, and discuss them.
Lastly, in Section~\ref{sec5}, we conclude our work and give future research directions.

\section{Related Work}
\label{sec2}

\subsection{Scientific Publications as a Datasource}

Related work in the technology forecasting domain generally uses technological texts, e.g., books or scholarly articles, as data sources.
Patents and scholarly articles have proven popular among researchers.
Therefore, in the following section, we critically analyze both and explain why we choose scholarly articles as our data source.

In~\cite{mikova_comparing_nodate}, the authors compare multiple data sources in their predictive value in the Green Energy domain.
They find that patents and academic articles contribute the most trends of the data sources analyzed.
Martino~\cite{martino_review_2003} classifies data sources along the innovation life cycle.
According to the author, basic and applied research can best be forecasted using scientific publications, while the development of specific technologies is best reflected in patents.

For these reasons, a large part of previous research uses patents as their primary data source.
A \emph{patent} is an intellectual property that gives specific rights to the patent holder.
Patents are a direct output of R\&D activities of companies and individuals and can thus be seen as a direct measure of technical innovation~\cite{von_wartburg_inventive_2005}.
Previous research has shown a direct link between patents and economic value~\cite{gambardella_value_2008}.
Furthermore, patents protect intellectual property from being copied.
Protecting shows direct economic interest in using the intellectual property protected by the patent~\cite{schmoch_indicators_1997}.
%Early technological forecasting work clustered similar patents into technology clusters counted the cluster sizes and forecasted them  (citation needed).
%After a link between individual patents and economic value and social value has been established, later work forecasted the value of individual patents (citation needed).
The direct link between patents and value and the fact that patent data is readily available in patent databases\footnote{\url{https://worldwide.espacenet.com/}} makes patents a popular choice for technological forecasting.

Establishing a direct link between economic value and basic research is more challenging.
So the main question is: why use scientific publications as a data source in technological forecasting?
The two main reasons are the scientific publications' strong influence on patents and the lead time between publications and patents~\cite{segev_analysis_2015}.

Previous work has shown strong links between publications and patents: the number of cited scientific articles in patents is increasing~\cite{narin_increasing_1997}, technology development is heavily influenced by public research~\cite{lo_scientific_2010}, increasingly patents get published by the same entities that publish scientific articles~\cite{schmoch_indicators_1997}, and the countries with the highest number of publications are also the countries with the highest number of patents~\cite{meyer_patent_2000}.
Therefore, we conclude that research activity and patenting activity are heavily linked.

Additionally, publications lead patent trends.
The authors of~\cite{segev_analysis_2015} aim to quantify forecasting lead times of different data sources by comparing patents and scientific articles to discover which source first signals technology trends.
They find that trends can be identified the earliest in scientific articles, four to five years in advance of patents, and nine years in advance of web searches and news articles.
The fact that trends in scientific articles predate trends in patents indicates that articles can be seen as a precursor to patents.

Therefore, by using scientific publications as our dataset, we are forecasting trends at an earlier stage than would be possible by using patents, and we are forecasting the same trends.
%Additionally, forecasting using scientific publications indicates increased patenting activity in the forecasted technology clusters in the future.

\subsection{Quantifying Technological Development}

We can segment previous approaches to quantifying technological development into quantitative~\cite{huang_technology_2022}~\cite{teixeira_technological_2012} and qualitative approaches~\cite{haegeman_quantitative_2013}.
Qualitative approaches rely on expert opinion to interpret the current technology landscape and forecast.
Relevant to this work are quantitative approaches, which we divide into two main clusters: direct measures and indirect measures.
We structure and justify the remainder of this subsection along these clusters.

Researchers use performance metrics of technologies to directly quantify technological progress, e.g., cost per unit, power output, efficiency, or energy consumption.
The improvement in a metric (e.g., decrease in cost per unit) is then equated with technological development.
In~\cite{way_empirically_2022}, the authors forecast the future cost of green energy solutions.
Depending on the future decrease, they present several scenarios, with lower costs leading to higher adoption.
In~\cite{managi_technological_2005}, the authors use drilling costs and oil and gas outputs to analyze the oil and gas industry change.
In~\cite{koh_functional_2006}, the authors create a database of performance metrics for information technologies and structure the metrics across three dimensions: storage, transformation, and transport.
In~\cite{magee_quantitative_2016}, the authors build on the previously presented work and forecast 28 technological domains using performance metrics on storage-, transformation-, and transport capabilities.
The main advantage of using performance metrics to forecast is the direct measurement of technological progress with measurable impact.
However, using direct measures also has its problems: gathering long-term performance data can be difficult, and only a subset of technologies where performance data exists can be forecasted.

To fill this gap, previous work also uses indirect measures to quantify technological progress.
Indirect measures are proxies for the indirect effects of technological progress.
Often, indirect measuremtn is done via bibliographic metrics, such as count data and citations.
The idea behind using these metrics is that technological progress is reflected in technological texts.
Thus, the underlying assumption is that progress can be proxied by the activity in technological texts measured by bibliometrics.

Previous citation analysis work mainly focuses on using patents as the data source.
These works mainly do citation analysis to attribute a measure of value to patents, e.g., highly cited patents indicating a valuable patent \cite{harhoff_citation_1999}.
In~\cite{choe_patent_2013}, the authors build a patent citation network to understand the structure of the organic photovoltaic domain.
In~\cite{daim_forecasting_2020}, the authors evaluate the blockchain and Internet of Things landscape using patent citation analysis, in~\cite{hung_technological_2014}, the authors build an evolution path of the lithium iron phosphate battery using a citation network built from academic articles.
In~\cite{li_monitoring_2020}, the authors build a citation network from academic articles and patents to conduct a gap analysis between them.

Besides citations, the number of scientific publications is an important metric to gauge activity in a technology field~\cite{hullmann_publications_2003}; thus, the earliest technology forecasting work analyzes document counts~\cite{rogers2014diffusion}, \cite{ayres_technological_1969}.
More recent examples are ~\cite{priestley_innovation_2020},~\cite{liu_forecasting_2010}, and~\cite{percia_david_security_2022}. 
The uptick of publications in a patent cluster indicates discoveries and technological progress.
Using document counts has two major drawbacks.
First, the count depends heavily on the underlying document taxonomy, i.e., how documents are assigned to technologies~\cite{von_wartburg_inventive_2005}.
An inaccurate taxonomy, where documents are misclassified, wrongly defines technologies and, thus, wrongly counts documents.
Second, counting documents equally does not reflect the fact that only a minority of documents accounts for the majority of the added value~\cite{gambardella_value_2008}.
A nominal count wrongly equals document contributions.

The choice of academic articles as our data source excludes individual article analysis to forecast technologies.
A link between individual scientific articles and economic value is difficult to establish since research predates technology diffusion by multiple years.
For example, a high citation count of academic articles does not necessarily translate to high economic value later~\cite{schmoch_tracing_1993}.
Thus, doing citation analysis on individual academic articles for technology forecasting has less utility.
Since analyzing document counts is an established technique in the technology forecasting domain, we use the count of published articles as a proxy for technological development.

\subsection{Previous Forecasting Approaches}

For our literature review, we divide related work focusing on forecasting indirect development metrics into two main approaches: model-based and data-driven approaches.
We structure the remaining paragraphs along these main approaches.

Technological development is the invention, innovation, and diffusion of new technologies, i.e., novel processes that produce outputs that were not feasible before~\cite{stavins_environmental_2002}.
An invention constitutes a scientifically or technologically new product or process, which may or may not develop into an innovation, which happens when an invention is commercialized~\cite{stavins_environmental_2002}.
A successful innovation becomes gradually more adopted by the market and society.
Model-based approaches generally assume a life cycle model that explains the technological development in the past, i.e., they are theory-driven.
Multiple model-based approaches exist; a popular one is the S-curve.
S-curves have a long history and appear in many different domains, e.g., Biology \cite{kingsland_refractory_1982}.

\begin{figure}[h]
    \centering
    \includegraphics[scale=0.5]{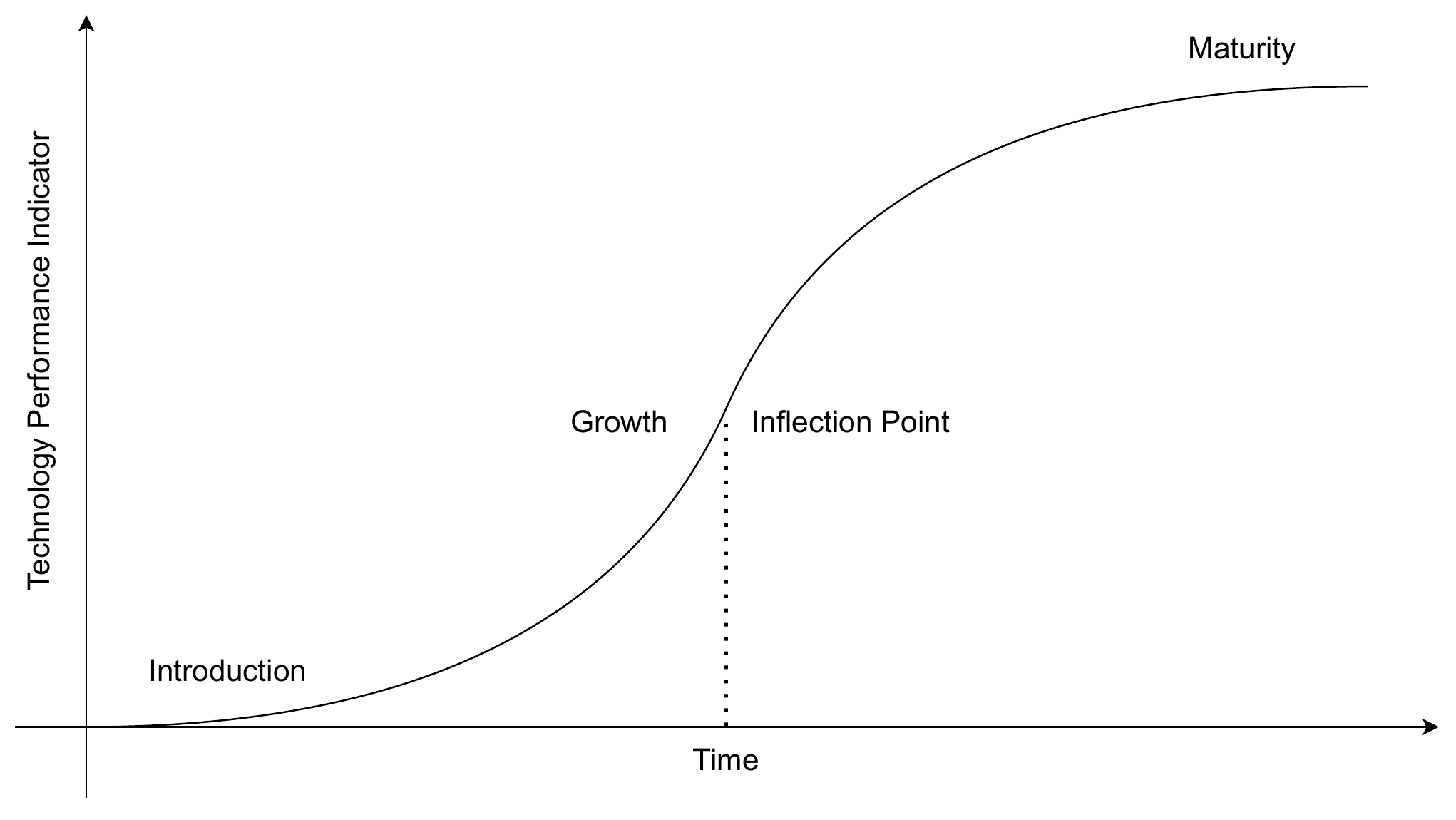}
    \caption{S-curve of technological development. The y-axis represents a \textit{technology-performance indicator} which captures the technological development over time (x-axis). Examples of performance indicators are adoption rate or the number of news articles published on a technology~\cite{percia_david_security_2022}.}
    \label{fig:single_sigmoid}
\end{figure}

S-curves model the life-cycle of technological development through three distinct phases \cite{priestley_innovation_2020}, \cite{lotfi_forecasting_2014}: the Introduction-, Growth-, and Maturity phases.
In the \emph{Introduction} phase, an innovation is new on the market and adopted by relatively few users.
It typically serves niche use cases, and changes to the technology happen slowly.
The technology is further developed during the \emph{Growth} phase and expands to mainstream use cases.
The technology has become an attractive target for investment and adoption; thus, the rate of technological development and adoption increase exponentially as the market punishes players for not using more efficacious and efficient technologies.
The technology approaches performance bottlenecks and other constraints in the \emph{Maturity} phase.
A technology stays at the Maturity phase until another innovation replaces it or new features allowing further development are added \cite{meyer_carrying_1999}.

S-curves have been found to appear in many different technological domains.
Meyer models the number of nuclear explosions, the population of Japan, the cumulative number of U.S. universities, and the adoption of electric generators using S-curves on time series data \cite{meyer_bi-logistic_1994}, \cite{meyer_carrying_1999}.
Kucharavy and De Guio model energy consumption and infrastructure development as S-curves \cite{kucharavy_logistic_2011}, Andersen \cite{andersen_hunt_1999} shows the prevalence of S-curves in 56 technological groups, ranging from chemicals to non-industrial fields.
Adamuthe et al., Priestly et al., and Percia et al. show the prevalence of S-curves in computer-science-related fields \cite{adamuthe_technology_nodate}, \cite{priestley_innovation_2020}, \cite{percia_david_security_2022}.

Apart from using S-curves to explain past behavior, related work also uses them to forecast.
In~\cite{kucharavy_logistic_2011}, the authors introduce a forecast segmentation.
They segment forecasts into short-, medium-, and long-term forecasts depending on how far into the future forecasts are to be made.
In~\cite{lotfi_forecasting_2014}, the authors focus on short-term forecasts (10-20\% of the existing data points), in~\cite{liu_forecasting_2010}, the authors fit S-curves on the number of patents in biped robot walking to find inflection points and saturation times and to do long-term forecasts on future trends for walking robots.
In~\cite{daim_forecasting_2006}, Daim et al. present a forecasting approach that uses S-curves fitted on the number of patent publications in three technologies.

The presented works show one major advantage of model-driven approaches: typically, models have interpretable parameters, e.g., in the case of S-curves, the inflection point where the growth rate starts to slow down.

One major disadvantage of model-based approaches is the limited usability of model-based forecasts.
When limited data is available (only the beginning of the S-curve), uncertainties in fits are high; thus, the utility of the fitted model is low.
If more data is available (S-curve is advanced a lot), the certainty of fit is high.
However, the utility for forecasting is also low, as most of the development has already happened~\cite{kucharavy_application_2011}.
This is limiting since accurate forecasts are especially useful in the early development phase of a new technology~\cite{priestley_innovation_2020},~\cite{lee_early_2018}.
Additionally, model-driven approaches assume an underlying growth model, e.g., the S-curve model in the case of the presented work.
Forecasting accuracy reduces significantly if the assumption is wrong.

The recent successes of Machine Learning have led to waves of adoption in different domains, technological forecasting included.
These machine learning approaches can capture complex non-linear relationships between input and output indicators and are thus especially strong at finding patterns in large amounts of data. 
They promise to tackle the disadvantages presented in the previous paragraph as they reduce the assumptions taken by finding patterns in the data in a supervised or unsupervised manner, i.e., they are data-driven.

As the direct economic value of single patents has been shown, most works employ data-driven methods to improve citation analysis.
Several works divide patents into emerging technologies and not emerging and train a model to classify patents into these two clusters correctly.
Whether a technology is emerging can be quantified using patent indicators such as forward and backward citations.
In~\cite{zhou_forecasting_2020}, \cite{kyebambe_forecasting_nodate}, and \cite{lee_early_2018} the authors segment patents into emerging and not emerging and then train a machine learning classifier to identify patents with future potential correctly.
In~\cite{choi_novel_2020}, the authors exploit the fact that the lifetime of a patent has been shown to be directly linked with economic value.
They use a machine learning model to predict the probability that a patent will survive until its maximum expiration date.
Other works use scholarly articles as their data source.
For example, in \cite{hassan_deep_2018} the authors classify important articles by their citation count. 
The work of Lee et al.~\cite{lee_pre-launch_2014} combines a model-based with a data-driven approach: the authors use a machine learning model to predict the model parameters of a life cycle model.

A first research gap we see is a direct comparison of model-based approaches to other forecasting approaches.
%However, their application does not always yield better results.
%For example, Dacrema et al. show that simple baseline methods can make more accurate recommendations than sophisticated machine learning algorithms.
%We see a similar trend in technological forecasting, where sophisticated machine learning models are developed to produce forecasts.
%A direct comparison between these two approaches has yet to be made.
Even though model-based approaches are frequently used to generate forecasts, they have not been compared to other forecasting methods.
Therefore, our work offers a critical analysis of their forecasting capabilities by comparing them to two other approaches.

While there is ample previous work on using machine learning approaches to forecast the value of individual patents, we find a second research gap in using machine learning approaches to forecast time series of technological development data.
As shown, model-based approaches produce forecasts making assumptions that can be wrong.
These assumptions can prove especially hard to make at the start of a time series, as insufficient time has passed for the time series to reveal its properties.
Machine learning has proven itself useful for forecasting in this setting.
Neural networks have achieved impressive results in sequence forecasting tasks, for example, in natural language processing (NLP) \cite{graves_generating_2014}, \cite{sutskever_sequence_2014}.
An architecture that has proven to be especially suited to sequential tasks is the recurrent neural network (RNN).
RNNs model the sequential nature of data explicitly in contrast to other architectures; for more details, see \cite{sherstinsky_fundamentals_2020}.
RNNs have recently been successfully applied to time series forecasting; see \cite{toubeau_deep_2019}, \cite{bandara_forecasting_2018}.
Especially interesting work is the one of Salinas et al. \cite{salinas_deepar_2019}, who present a probabilistic sequence to sequence (seq-2-seq) time series model that forecasts the demand of products on Amazon using RNNs.
Their approach produces more accurate forecasts than baseline methods with interpretable confidence intervals for their forecasts.
However, the model has yet to be applied to technological development data.

To fill the first research gap, we compare the S-curve's forecasting performance to two forecasting approaches: ARIMA models and the data-driven approach we develop in this work.
We test S-curves as a model-based forecasting approach because of their widespread presence in technological development data and ARIMA models because of their widespread and easy use.

To address the second research gap, we develop a data-driven forecasting approach based on the work of Salinas et al.~\cite{salinas_deepar_2019} to predict the number of scholarly publications in technological fields.
We focus on applying their model architecture to technological development data from \verb|arXiv| data to forecast.

\section{Data and Methods}
\label{sec3}

\noindent
In this section, we first introduce the dataset we use in this work and how we construct an empirical variable acting as the technology performance indicator (TPI) we are forecasting.
Afterward, we explain how we generate technology forecasts using S-curves, ARIMA models, and Salinas' time series forecasting approach and how we evaluate their performance.

\subsection{Data}
In this work, we use \verb|arXiv| as our dataset, an open-access distribution service created in 1991 of more than two million scholarly articles related to more than \num{170} technical domains.
It is openly available for download and comprises more than \num{3} TB of \verb|.pdf| files of scientific texts and their associated metadata~\cite{arxiv_dataset}.
We use the e-prints uploaded to the \verb|arXiv| as the data source because they represent an exhaustive body of knowledge in many technical fields.

\verb|arXiv| organizes publications into eight main categories and provides a definition of its taxonomy\footnote{\url{https://arxiv.org/category_taxonomy}}.
The main categories are Computer Science, Economics, Electrical Engineering and Systems Science, Mathematics, Physics, Quantitative Biology, Quantitative Finance, and Statistics.
These main categories further break down into 170+ subcategories.
The taxonomy of Computer Science fields is based on the ACM Computing Classification System and distinguishes between 40 different subfields.
The taxonomy of the other main categories is explicitly defined by \verb|arXiv| on their website.
Researchers classify their e-prints when uploading them.
Moderators then check the classification.
We assume the classification to be robust because the taxonomy was reached through consensus by a board of experts, researchers have the incentive to classify their research correctly, and moderators check the classification~\cite{percia_david_security_2022}.
We assume each subcategory on \verb|arXiv| to form an exclusive and exhaustive body of knowledge representing different technologies.
Therefore, in the subsequent text, when referring to a subcategory, we refer to the technologies it represents.

\begin{figure}[h]
    \centering
    \includegraphics[scale=0.3]{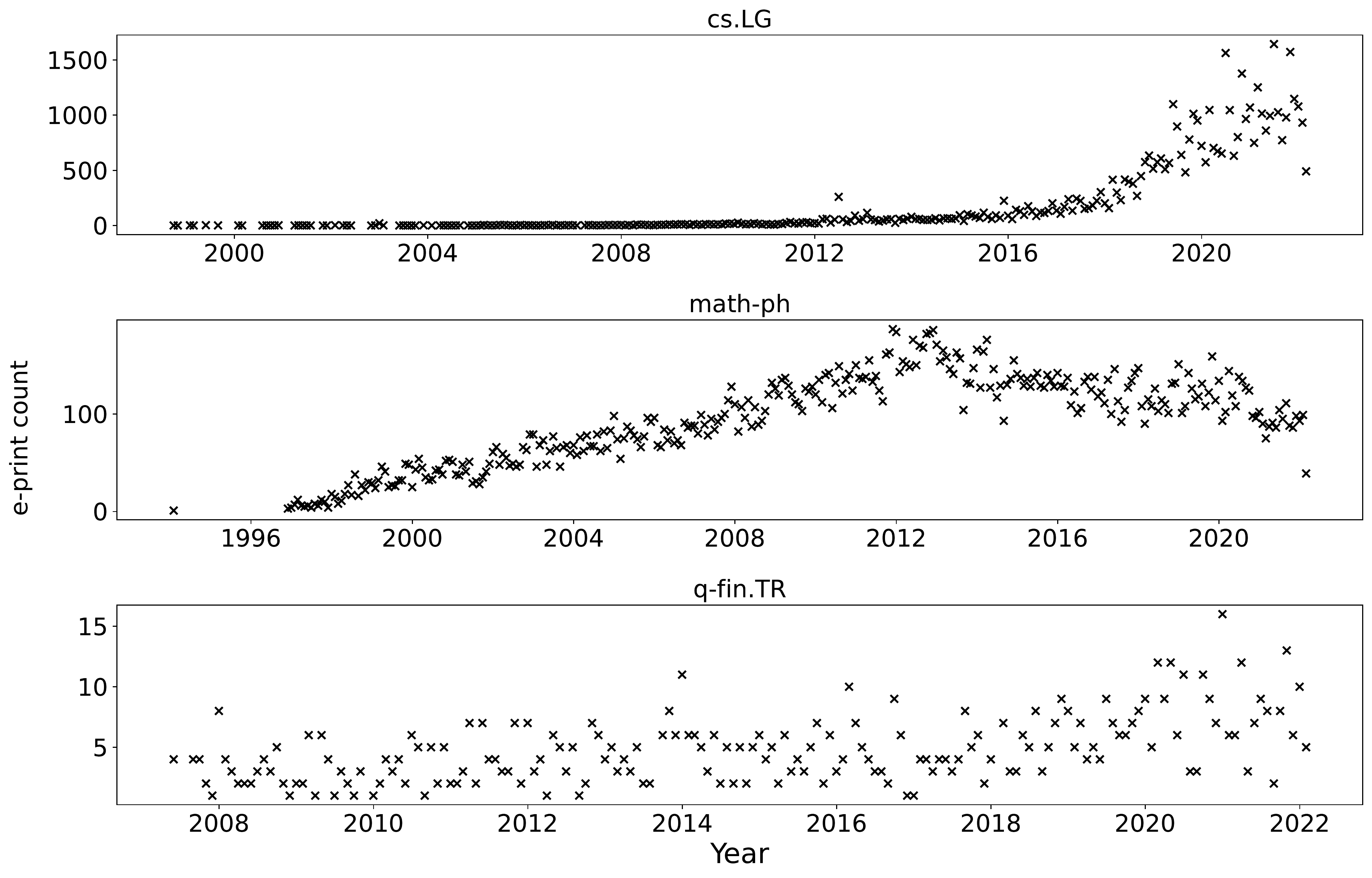}
    \caption{Monthly e-print count for selected subcategories on arxiv. cs.LG, math-ph, and q-fin.TR designate the Computer Science Machine Learning, the Mathematical Physics, and the Quantitative Finance Trading and Market Microstructure subcategories, respectively. The three subcategories show different development patterns and are not stationary.}
    \label{fig4}
\end{figure}

The metric we use to quantify technological development is the number of e-print uploads to \verb|arXiv|.
This number is a proxy for the development of the technologies described in the e-prints.
We calculate this metric monthly to balance the velocity and the amount of data (yearly aggregation yields too few data points and aggregates too much, and daily aggregation yields a time series with high variance).
This produces one time-series $z_i$ of the number of e-prints uploaded per month for each subcategory on \verb|arXiv|, cf. Figure~\ref{fig4}.
We exclude some subcategories, as their time series are too short, yielding 148 time series.

\subsection{Methods}

In the following subsection, we explain how we develop a model-driven and a data-driven technology forecasting approach and how we compare their performance.
We show how we produce forecasts on the arXiv data using S-curves as our model-driven approach and ARIMA as our baseline.
Additionally, we apply Salinas' et al.~\cite{salinas_deepar_2019} model to technology forecasting and evaluate its capabilities on technology forecasting data.
Therefore, we detail how we implement their model, train it, and forecast with it.

\subsubsection{S-curve}

\begin{figure}[!h]
    \centering
    \includegraphics[scale=0.6]{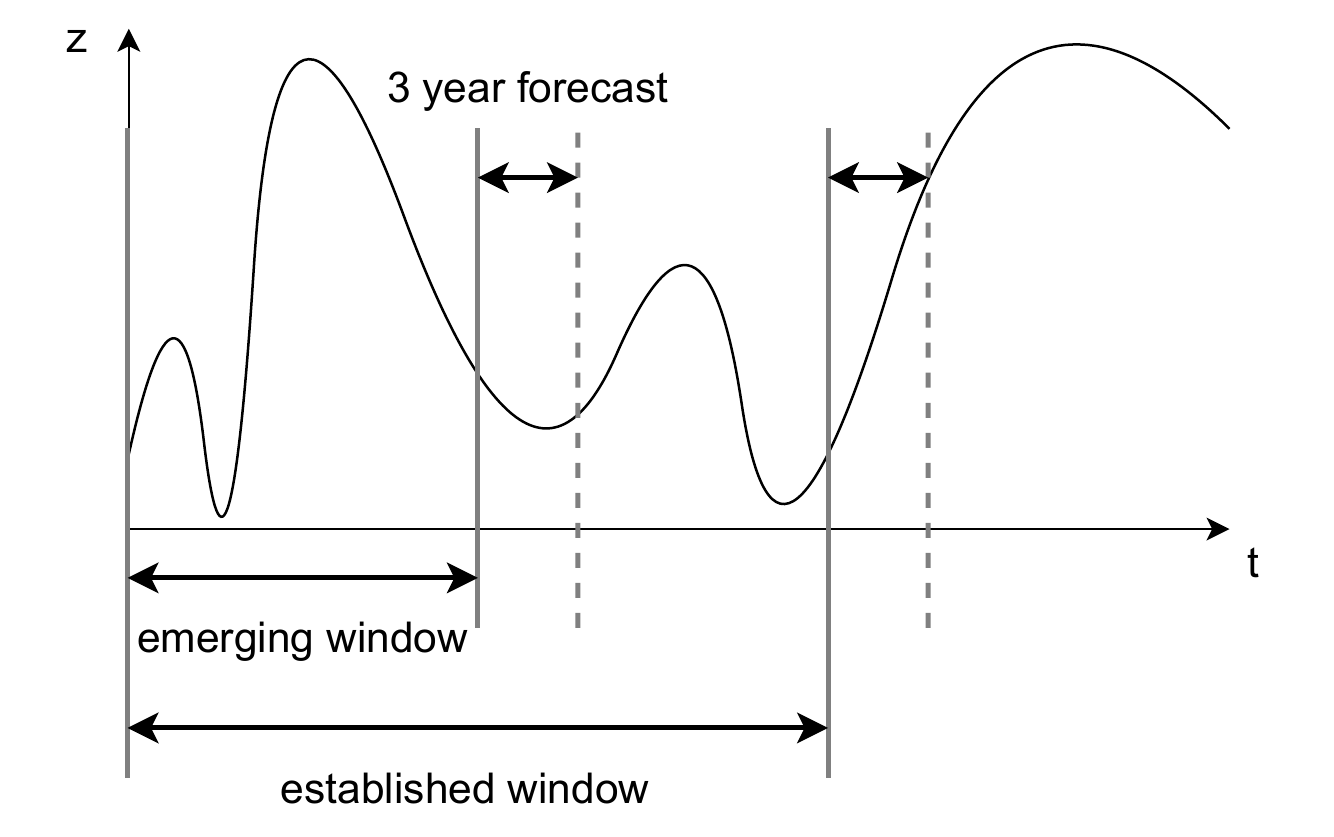}
    \caption{We evaluate the forecasting approaches at two stages in the subcategory life cycle. The emerging technology windows contain the first third of a given time series and are then used by a given forecasting approach to forecast the subsequent three years. The established technology windows consist of two-thirds of a time series. Applying this transformation to the 148 time series in our dataset yields 296 forecasting windows.}
    \label{fig:window_types}
\end{figure}

To produce S-curve forecasts as practitioners would do, we create two types of forecasting windows to simulate technologies in different stages in their life cycle, cf. Figure~\ref{fig:window_types}.
The first window type simulates technologies early in their life cycle, where only little data is available.
We refer to these windows as \emph{emerging technology windows} or simply \emph{emerging windows}.
These windows contain the first third of a given time series.
The second window type simulates technologies later in their life cycle.
We refer to these windows as \emph{established technology windows}, or \emph{established windows}.
The established windows contain the first two-thirds of a time series.
Applying this procedure to the 148 time series in our dataset, we get 296 windows, one early life cycle, and one late life cycle window for each time series.
For both types of windows, we produce three-year forecasts.
We fit the S-curve formulation onto the available data for each window using a non-linear fitting procedure.
The fitting requires hyperparameters to constrain the parameters of the S-curve.
We conduct a grid search for each window to find the hyperparameters that yield the most accurate fit.
After calculating an S-curve fit on every window, we produce forecasts by calculating the number of uploaded papers for a technology as given by the S-curve and compare that to the observed number of uploads by calculating the RMSE and MAPE.
In the following, we refer to forecasting using S-curves as the FIT approach.

\subsubsection{ARIMA}

To introduce a baseline, we directly compare the FIT forecasts to forecasts produced by ARIMA models.
ARIMA models are state-of-the-art and widely used to forecast time series.
We use the same windows we created for the FIT forecasts.
All models are ARIMA(1, 0, 1) models; any other configuration caused convergence problems due to some time series in our dataset being too short for larger orders.
However, that does not pose problems, as we use the ARIMA models only as a baseline to compare against.
%On each window, we run a grid search to find the $p, d, q$-order that produces the most accurate fit on the window.
Here again, we produce a three-year forecast and calculate RMSE and MAPE.

\subsubsection{RNN}

Forecasting using S-curves and ARIMA models is fundamentally different from forecasting with the RNN model.
With S-curves and ARIMA models, we fit a different forecasting model to every window, whereas the RNN model trains on all time series simultaneously to create one forecasting model.
As such, we have to present the RNN forecasting approach differently.
To present it, we distinguish between implementation, training, and evaluation.
First, we present how we implement Salinas' model to forecast technological development on our arXiv dataset.
Second, we go into detail about how we train the model.
Third and last, we explain how we evaluate the model and compare its performance to the FIT and ARIMA approaches.

\begin{figure}
\centering
\begin{subfigure}{.48\textwidth}
    \centering
    \includegraphics[width=\linewidth]{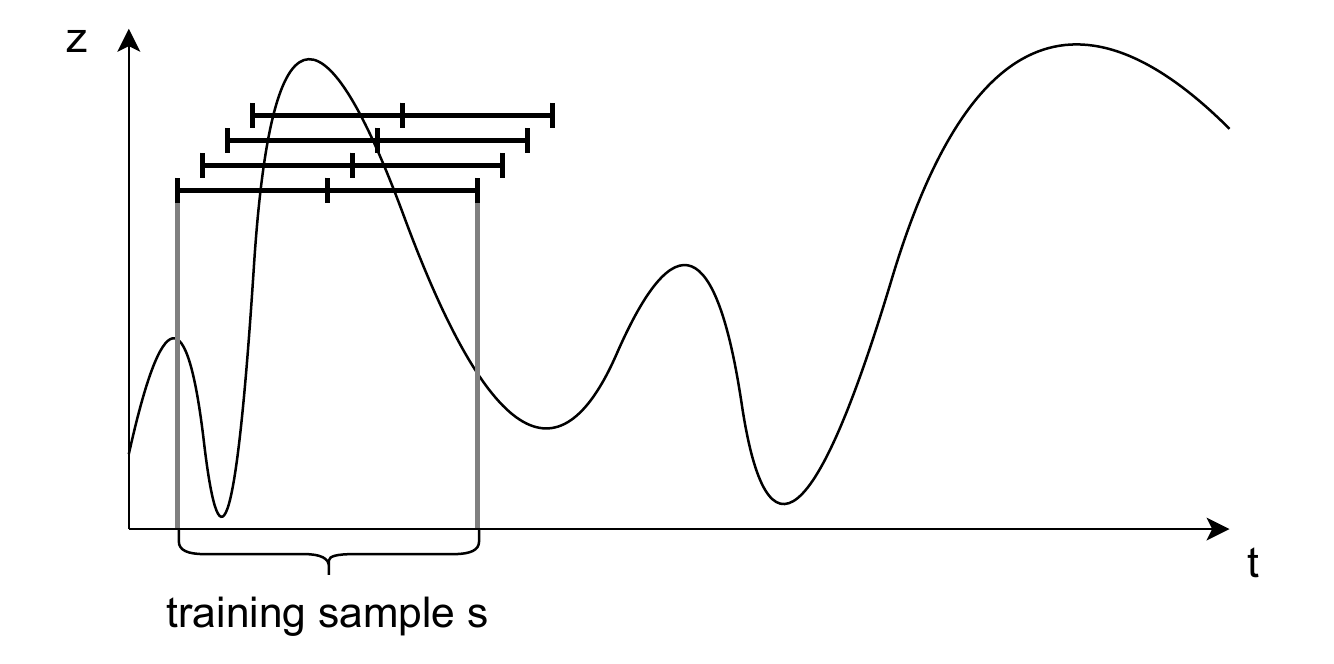}
    \caption{Example time-series from which we create training samples $s$.}
    \label{fig7:1}
\end{subfigure}
\hfill
\begin{subfigure}{.48\textwidth}
    \centering
    \includegraphics[width=\linewidth]{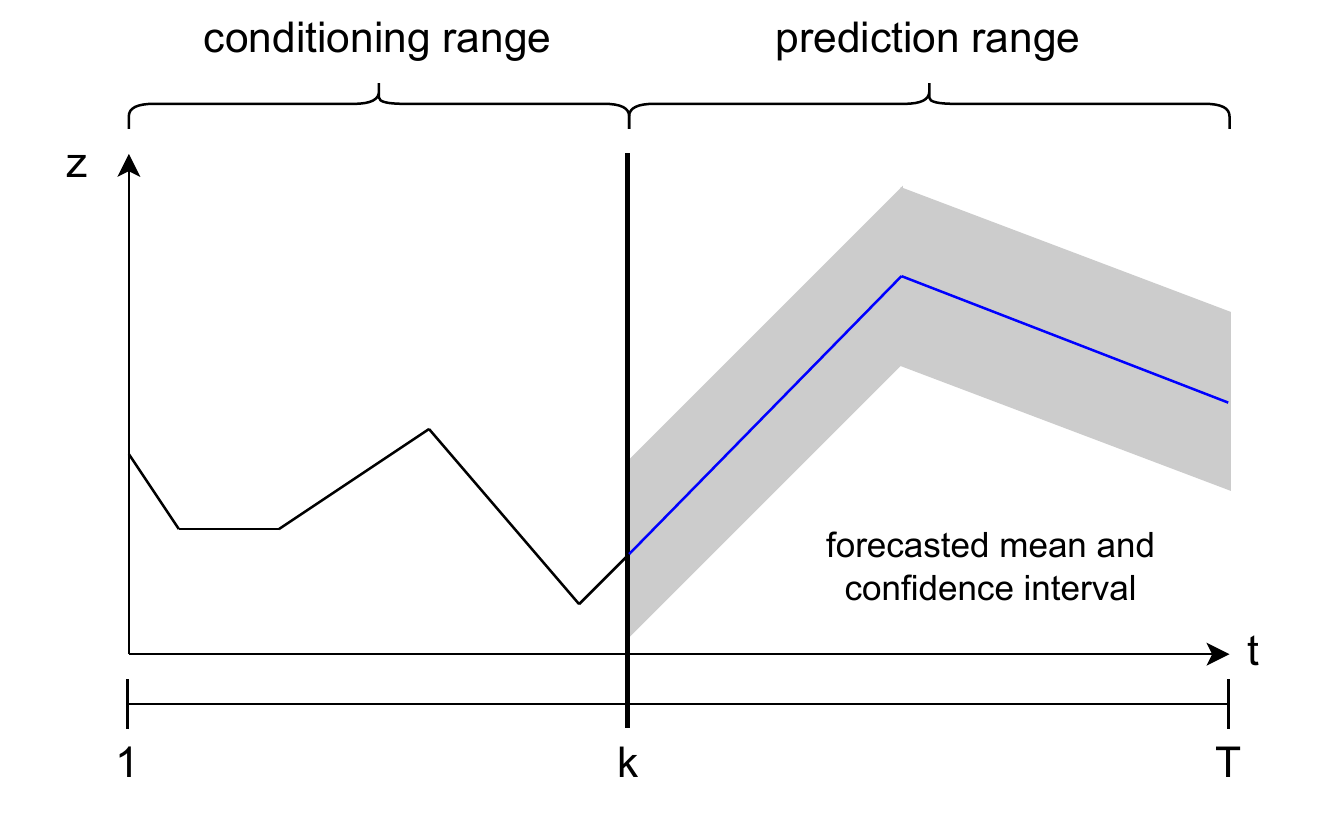}
    \caption{Setup of one training sample $s$ of the time-series $z$ taken from Salinas et al.~\cite{salinas_deepar_2019}.}
    \label{fig7:2}
    \end{subfigure}
\caption{Dataset augmentation technique we use to create training samples~\cite{salinas_deepar_2019}. On the left, we show how we generate multiple training samples from one time series. We slide a fixed-length window over the time series to create the training samples. Thus, we can create multiple training samples from one time series. In the example above, we create four training samples. The right graph shows the structure of one training sample. The vertical line splits the window into the conditioning range ($1, \dots, k$) and the prediction range ($k + 1, \dots, T$). 
%The parameters $\Theta_t$ of the probability distribution predicted by the model determine the position of the mean and the confidence interval.
}
\label{fig7}
\end{figure}

Sequence-to-sequence (seq-2-seq) models
%(cf. Appendix for a more detailed description of the architecture), 
 like the one Salinas et al. developed, are powerful at modeling the sequential nature of data.
They map an input sequence of arbitrary length to an output sequence of arbitrary length.
These models consist of two parts: an encoder and a decoder.
The encoder encodes the input time series $z_i$ into a hidden state $h_i$.
The decoder decodes that hidden state into the output.
The output depends on the hidden state; thus, on the input, i.e., the decoder is conditioned on the input $z_i$.
Both the encoder and decoder are trained in parallel to map an input to the desired output as accurately as possible.
We augment the dataset by generating multiple training samples from each time series, cf. Figure~\ref{fig7:1}.
We divide a training sample into a \emph{conditioning} range and \emph{prediction} range.
%The conditioning range is the input to the model for which we want to forecast the development in the prediction range.
During the training, the model forecasts the values in the prediction range using the conditioning range as input.
The forecasted values are then compared to the observed values in the prediction range, and the error is backpropagated to train the model weights.
During testing, only values in the conditioning range are available.
To ensure comparability with the previous two forecasting approaches, we set the length of the prediction range to 36 months, i.e., three years.
Choosing the length of the conditioning range is a trade-off between the number of training samples and forecasting accuracy.
Increasing the conditioning range increases forecasting accuracy and decreases the number of training samples.
We set the length of the conditioning range to 36 months, or three years, to strike a balanced trade-off.
In total, each training window is six years long.

To generate a training sample, we slide a window of fixed-length $l$ from the beginning $t_1$ of the time series to the end, where each starting date marks a new training sample.
This technique generates multiple training samples of the same length $l$ from each time series.
Our data set comprises \num{148} distinct time series, from which we generate around \num{13000} training samples.

\begin{figure}[!h]
    \centering
    \includegraphics[scale=0.3]{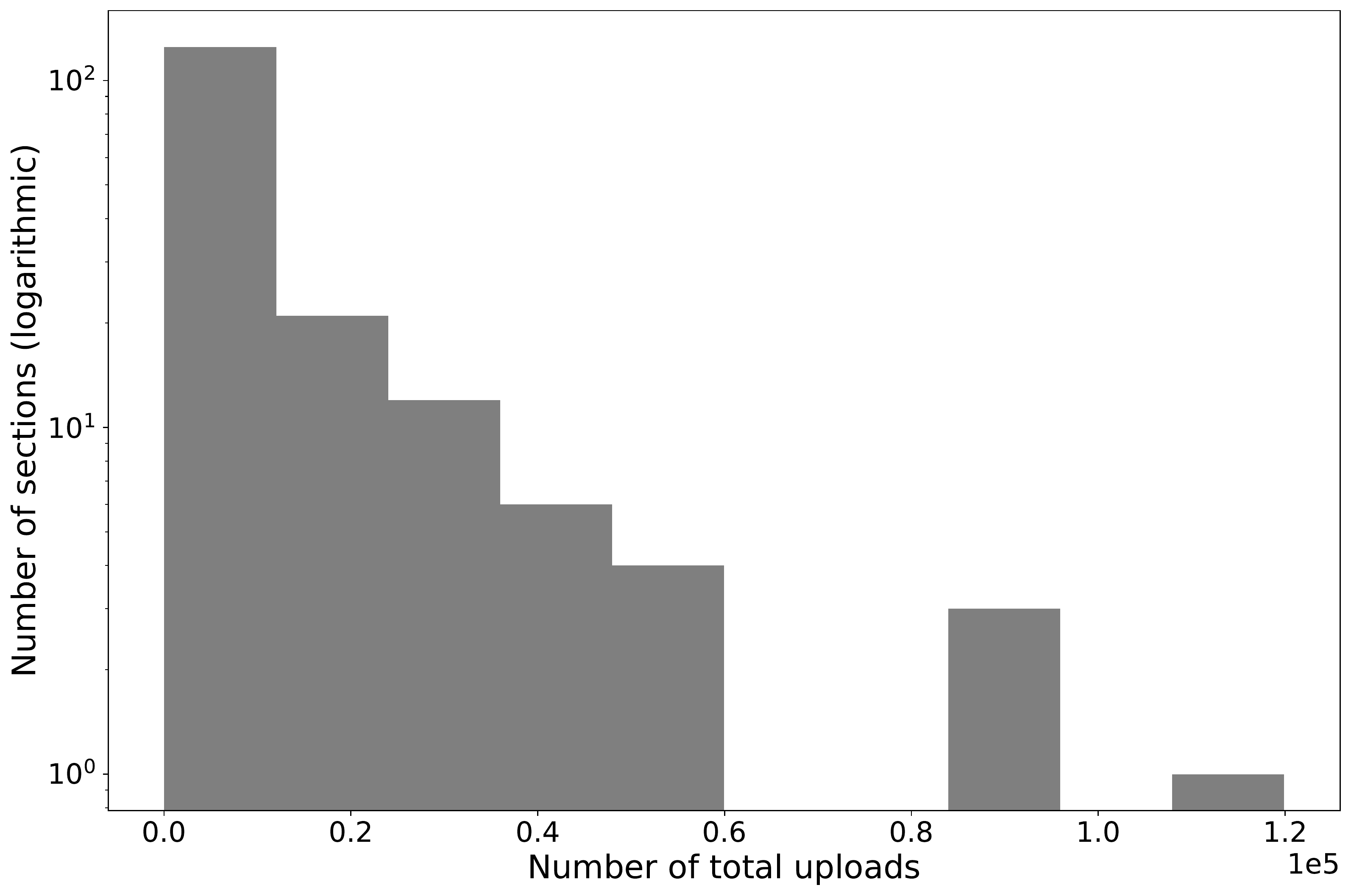}
    \caption{Histogram of the number of uploads per section. A majority of subcategories get few uploads, while a minority of subcategories are very popular, i.e., distribution of uploads in subcategories on arXiv follows a power law. Therefore, we scale the number of uploaded e-prints to render the training of our model more robust.}
    \label{fig9}
\end{figure}
The number of e-prints uploaded on \verb|arXiv| varies from subcategory to subcategory; thus, each time series has a different scale.
Figure~\ref{fig9} shows the unequal distribution of papers between the subcategories on \verb|arXiv|.
Without scaling, the model would have to learn to scale appropriately on top of forecasting, yielding worse accuracy and convergence.
In these cases, it is standard practice to scale the affected variables.
In this work, we scale a training sample by the mean value of the time-series $z_i$ in the conditioning range~\cite{salinas_deepar_2019}: $z_{i,scaled} = \frac{z_i}{v_i}$ with $v_i = 1+ \frac{1}{k}\sum_{t=1}^{k}z_{i, t}$.

To predict the development of a subcategory from beginning to the end, we split the data set across the subcategories into training-, validation-, and test set.
We randomly choose 10\% of sections to be in the validation- and test set, all training samples generated from these subcategories are never seen during training.
If we took the beginning of each time series as our training set and the end as our validation- and test set, the training and evaluation would be biased: during training, the model would never see samples from the end of a time series.
Conversely, during testing, the models would never be evaluated for predicting the beginning of a time series.
Therefore, by choosing whole subcategories as the test set, we can evaluate the model's forecasting capabilities along the whole life cycle of technologies.

We conduct a hyperparameter search to find the best training parameters and the most accurate model configuration.
We vary the models across two dimensions: the model type (RNN vs. LSTM), and the model complexity (number of hidden nodes).
It is standard practice in Deep Learning to vary the model complexity and model type to find the simplest model structure that can still forecast accurately.
We determined the training parameters in a previous hyperparameter search.
To train the model, we use a batch size of \num{64}, a learning rate of \num{0.001}, and \num{750} training epochs.

For the evaluation of the RNN approach, we use the forecasting windows shown in Figure~\ref{fig:window_types}.
Reusing the same windows used to evaluate the two previous forecasting approaches permits an unbiased comparison.
However, we cannot use all evaluation windows because they contain the RNN training set windows.
We only use forecasting windows from subcategories in the RNN test set to avoid false accuracy metrics to evaluate the RNN's accuracy.

\section{Results and Discussion}
\label{sec4}

%\begin{table}[]
%\centering
% Table 1
%\begin{minipage}[t]{0.48\linewidth}
%\centering
%\begin{tabular}{lll}
%Method & RMSE mean/median & MAPE mean/median \\ \hline
%FIT    & 64.3/9.0 & 984.8/\textbf{36.3}\\
%ARIMA  & \textbf{18.7}/\textbf{8.2} & \textbf{49.6}/38.2
%\end{tabular}
%\label{tab:fit_vs_arima}
%\caption{Forecasting accuracy for the FIT and ARIMA methods on all 148 technologies.}
%\end{minipage}\hfill%
% Table 2
%\begin{minipage}[t]{0.48\linewidth}
%\centering
%\begin{tabular}{lll}
%Method & RMSE mean/median & MAPE mean/median \\ \hline
%FIT    & 249.3/11.6 & 2955.8/36.2 \\
%ARIMA  & 19.0/\textbf{8.8}  & 55.0/33.7   \\
%RNN    & \textbf{14.1}/9.4  & \textbf{41.5}/\textbf{32.5}
%\end{tabular}
%\label{tab:fit_vs_arima_vs_rnn}
%\caption{Forecasting accuracy for the FIT, ARIMA, and RNN methods on the test technologies.}
%\end{minipage}
%\caption{The mean and median forecasting accuracy for the FIT, ARIMA, and RNN methods. We include mean and median as the forecasting errors are right-skewed. We highlight the lowest score in each column bold.}
%\end{table}

%\begin{adjustbox}{width={\textwidth},totalheight={\textheight},keepaspectratio}
\begin{table}[!htb]
    \begin{subtable}{.47\textwidth}
    \centering
    \begin{adjustbox}{width=1\textwidth}
    \begin{tabular}{lll}
        Method & RMSE mean/median & MAPE mean/median \\ \hline
        FIT    & 64.3/9.0 & 984.8/\textbf{36.3}\\
        ARIMA  & \textbf{18.7}/\textbf{8.2} & \textbf{49.6}/38.2
    \end{tabular}
    \end{adjustbox}
    \caption{Forecasting accuracy for the FIT and ARIMA methods on all 148 technologies. Both approaches forecast with similar accuracy in the median; however, due to outlier predictions the mean accuracy metrics for the FIT approaches are significantly larger compared to the ARIMA approach.}
    \label{tab:fit_vs_arima}
    \end{subtable}%
    \hfill
    \begin{subtable}{.47\textwidth}
    \centering
    \begin{adjustbox}{width=1\textwidth}
    \begin{tabular}{lll}
        Method & RMSE mean/median & MAPE mean/median \\ \hline
        FIT    & 249.3/11.6 & 2955.8/36.2 \\
        ARIMA  & 19.0/\textbf{8.8}  & 55.0/33.7   \\
        RNN    & \textbf{14.1}/9.4  & \textbf{41.5}/\textbf{32.5}    
    \end{tabular}
    \end{adjustbox}
    \caption{Forecasting accuracy for the FIT, ARIMA, and RNN methods on the test technologies. The RNN approach produces the most accurate forecasts.}
    \label{tab:fit_vs_arima_vs_rnn}
    \end{subtable}%
    \caption{The mean and median forecasting accuracy for the FIT, ARIMA, and RNN methods. We include mean and median as the forecasting errors are right-skewed and highlight the lowest score in each column in bold.}
\end{table}
%\end{adjustbox}

% performance fit vs arima
Comparing the performance of the FIT and ARIMA models, cf. Table~\ref{tab:fit_vs_arima}, we see that the FIT approach exhibits one magnitude higher mean error metrics and that the error distributions are right-skewed.
The FIT approach has a mean MAPE of 985\% versus the ARIMA approach, which has a mean MAPE of 50\%.
However, the median error metrics are similar (the largest difference is 1.9\%), around 37\%.
We trace this behavior to outlier forecasts of the FIT approach.
On a minority of windows, the FIT approach forecasts very poorly.

\begin{figure}[h]
\centering
\includegraphics[width=0.7\textwidth]{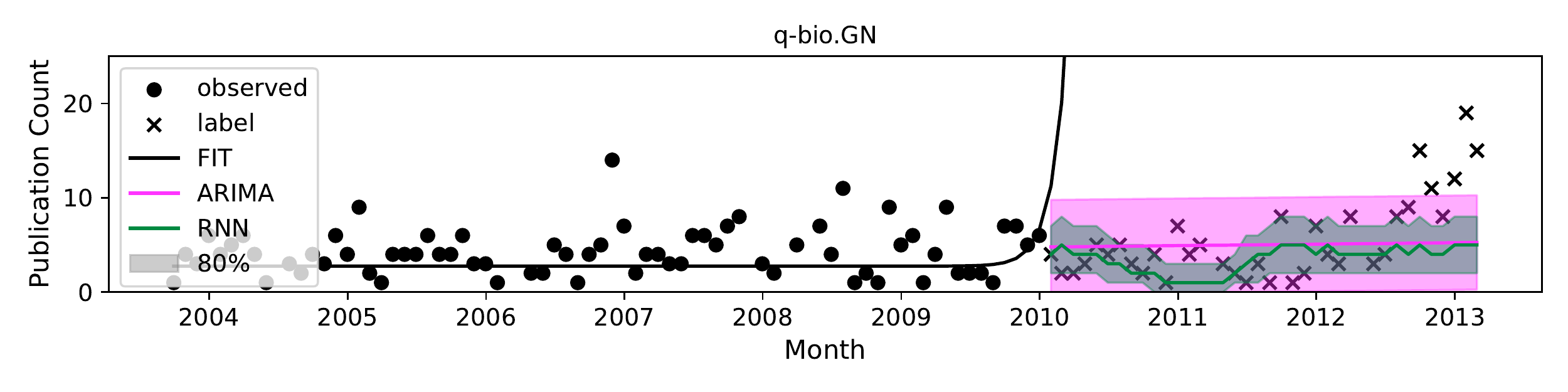}
\caption{Qualitative comparison of forecasts produced by each approach. While the forecasts of the ARIMA and RNN approaches are close, the FIT approach forecasts a false growth spurt. We attribute this behavior to the fact that the fitting algorithm needs to fit a growth spurt to data where no real growth has occurred. The fitting algorithm exploits this to reduce the fitting error by putting an exponential growth phase at the end of the observed data.}
\label{fig:forecast_comparison}
\end{figure}

% performance fit
The poor performance of the FIT approach is caused by its tendency to forecast exponential growth phases for windows where the S-curve's growth phase has yet to happen.
Looking at Figure~\ref{fig:forecast_comparison}, we see that the fitting procedure fits the S-curve well to the observed data.
However, since no growth happened, the fitting procedure can exploit that fact by putting the beginning of an exponential growth phase at the end of the observed data.
This reduces the in-sample fit error but produces erroneous forecasts.

The tendency of the FIT approach to forecast erroneous growth spurts is due to the assumption that the development follows an S-curve pattern.
The S-curve formulation includes a growth spurt.
If a given observed life cycle does not exhibit a growth spurt, an S-curve will not accurately describe the development.
The main problem is that at the beginning of a technology's life cycle, practitioners cannot know which life cycle model a technology will follow.
This shows the central crux of using life cycle models: to produce useful forecasts, practitioners need to forecast technologies as early as possible without having enough information to make correct assumptions.

% performance arima
%The mean RMSE and MAPE for the FIT approach are a magnitude higher than for the ARIMA approach.
%However, the median metrics are similar.
%We trace this behavior to outlier forecasts of the FIT approach.
%On a minority of windows the FIT approach forecasts very poorly.
%Comparing the forecasts using Figure~\ref{fig:forecast_comparison}, we see that even though the ARIMA approach does not forecast false growth spurts like the FIT approach, it misses the growing trend visible in subcategory cs.SD.
%This is not surprising, as the ARIMA simply does not allow for changing trends like this to be forecasted. 

% performance fit vs arima vs rnn
Comparing the RNN approach to the previous two approaches, cf. Table~\ref{tab:fit_vs_arima_vs_rnn}, we see that the RNN approach produces the most accurate forecasts on average.
Its MAPE is at 41.5\% on average, 13.5\% lower than the next best result, the ARIMA approach.
The ARIMA model has the second-best mean MAPE of 55\%.
However, its median MAPE at 33.7\% is only 1.2\% higher than the RNN's median MAPE, indicating a more pronounced right-skew.
Again all error distributions are right-skewed, with the FIT approach exhibiting the lowest accuracy and the heaviest right-skew. 
Its mean MAPE of 2955.8\% is two magnitudes higher, while its median MAPE at 36.2\% is only 2.5\% higher compared to the next best result.
%Comparing the results qualitatively using Figure~\ref{}, we similar results for RNN and ARIMA.
%While both are able to forecast trends, they are not able to capture sudden changes in trends, as seen in subcategory cs.SD.

\begin{figure}[h]
\centering
\includegraphics[width=0.7\textwidth]{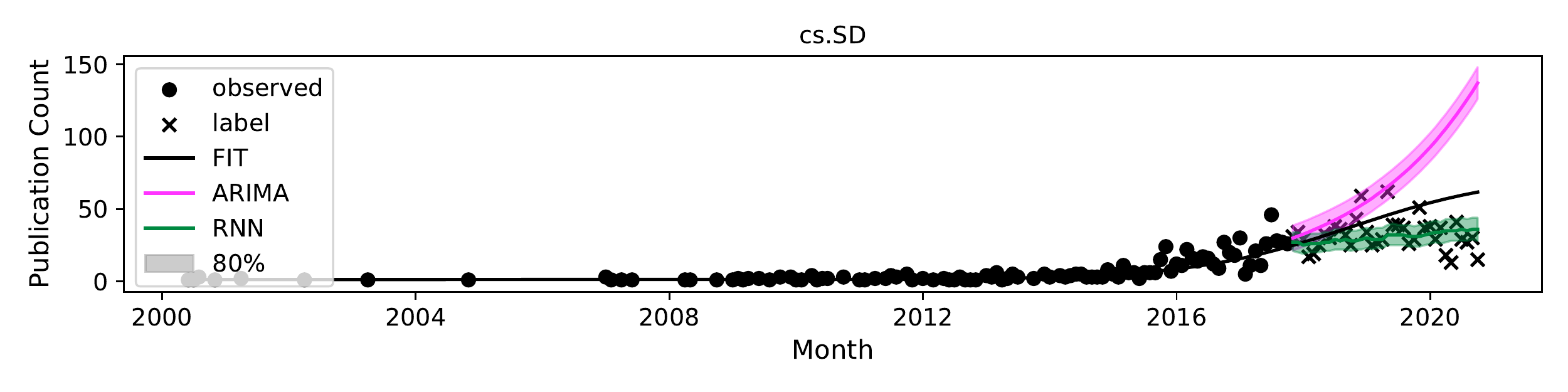}
\caption{Qualitative comparison of forecasts produced by each approach. The ARIMA approach predicts the accelerating trend to continue, while the RNN approach captures the reversing trend.}
\label{fig:forecast_arima_vs_rnn}
\end{figure}

% performance ARIMA vs RNN
The biggest difference in MAPE between ARIMA and the RNN approaches comes from windows where the ARIMA model predicts accelerating growth.
Although growth is slowing, the RNN predicts the slowing growth correctly, cf. Figure~\ref{fig:forecast_arima_vs_rnn}.
The ARIMA model's formulation cannot capture the reversing trend in these cases.
It predicts an accelerating growth if the growth was accelerating in-sample.

\begin{figure}[h]
\centering
\includegraphics[width=0.7\textwidth]{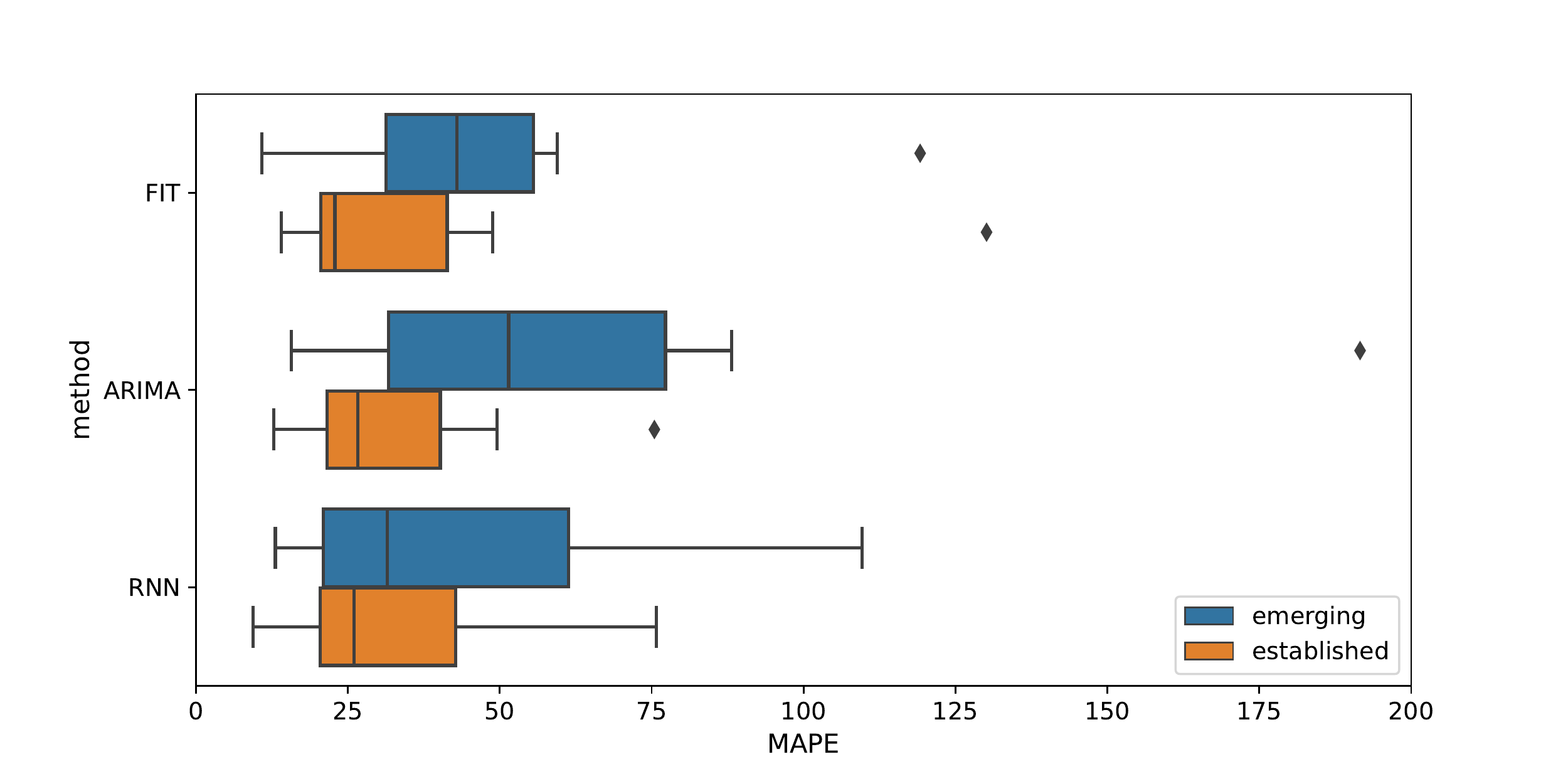}
\caption{Boxplots visualizing the MAPE accuracy distribution for each forecasting approach drilled down by window type. All models predict established windows with similar accuracy; the RNN approach predicts emerging windows more accurately than any other approach.}
\label{fig:boxplot}
\end{figure}

% performance on different performance
All forecasting methods forecast the established windows more accurately than the emerging windows, cf. Figure~\ref{fig:boxplot} and Table~\ref{tab:early_vs_late_windows}.
The median and mean MAPEs for the established windows are, without exception, lower than the emerging window MAPEs for every method.
The emerging window median MAPEs are 20.1\%, 24.8\%, and 6.7\% higher than the established window MAPEs for the FIT, ARIMA, and RNN approaches, respectively.
However, we observe that the RNN forecasts the emerging windows the most accurately of all approaches with a mean MAPE of 42.6\%.
The ARIMA approach achieves the next best MAPE of 60.2\%.
Therefore, we conclude that the performance advantage of the RNN approach comes from its ability to forecast emerging technologies more accurately than the other approaches.
\begin{table}[!htb]
    \begin{subtable}{.47\linewidth}
    \centering
    \begin{adjustbox}{width=1\textwidth}
    \begin{tabular}{lll}
        Method & RMSE mean/median & MAPE mean/median \\ \hline
        FIT    & 477.9/11.9 & 5876.4/42.9 \\
        ARIMA  & 12.8/7.7 & 60.2/51.5  \\
        RNN    & \textbf{9.4}/\textbf{6.2}  & \textbf{42.6}/\textbf{31.6} 
    \end{tabular}
    \end{adjustbox}
    \caption{Forecasting accuracies on the emerging windows.}
    \label{tab:early_windows}
    \end{subtable}%
    \hfill
    \begin{subtable}{.47\linewidth}
    \centering
    \begin{adjustbox}{width=1\textwidth}
    \begin{tabular}{lll}
        Method & RMSE mean/median & MAPE mean/median \\ \hline
        FIT    & 20.7/11.4 & \textbf{35.2}/\textbf{22.9} \\
        ARIMA  & 25.1/\textbf{9.0}  & 49.7/26.7   \\
        RNN    & \textbf{18.6}/10.6  & 38.95/24.9    
    \end{tabular}
    \end{adjustbox}
    \caption{Forecasting accuracies on the established windows.}
    \label{tab:late_windows}
    \end{subtable}%
    \caption{The mean and median forecasting accuracy for the FIT, ARIMA, and RNN methods drilled down by the window type. Emerging windows represent the first third of a technologies life cycle, and established windows represent the first two-thirds. While all models predict established windows with similar accuracy, the RNN approach predicts emerging windows (which are important for practitioners) 18\% more accurately on average than the next best approach.}
    \label{tab:early_vs_late_windows}
\end{table}
% Boxplot

\section{Conclusion}
\label{sec5}

Even though S-curves and other technological development models have been used extensively to forecast, our literature review revealed a research gap in their direct evaluation against other established forecasting methods.
Additionally, model-based approaches suffer from one major pitfall: their forecasts require sufficient data to produce accurate forecasts.
However, the most benefit can be had when forecasting emerging technologies, for which sufficient data is usually not available.
We identify a second research gap in applying modern machine learning-based, data-driven forecasting approaches to tackle these issues.

To address the two presented research gaps, we develop a practitioners guide on producing forecasts using S-curves and benchmark their performance against ARIMA forecasts.
Additionally, we apply a probabilistic time series forecasting approach developed by Salinas et al.~\cite{salinas_deepar_2019} to technological development data and compare its performance to the previous two forecasting approaches.

We find that the S-curve and ARIMA approaches have comparable median forecast accuracies.
However, in a minority of samples, the S-curve forecasts erroneous growth spurts on new technologies, which increases the mean forecasting errors by two magnitudes over the ARIMA approach.
The RNN approach forecasts established windows with the same accuracy as the two previous approaches but emerging windows with a mean MAPE 18\% lower than the next best result.

Our findings have a significant implication for practitioners looking to employ S-curves or other model-based approaches to forecast technologies.
S-curves have been proven to be useful in explaining past development.
However, our work casts doubts on their utility in producing useful and accurate forecasts.
Practitioners using a quantitative forecasting approach should consider using a simple ARIMA model to forecast over S-curves, as it has proven to be more accurate for emerging technology prediction and as accurate as S-curve predictions for established technologies.

The RNN approach we present in this work is promising thanks to its accurate forecasts, but it is more complex to implement.
We advise practitioners looking for quick estimating forecasts to employ a simple ARIMA model.
Practitioners seeking more accurate forecasts, especially on emerging technologies, should opt for the RNN approach.

We see further venues to improve the forecasting accuracy of the RNN approach by finding and testing other predictive features.
In our work, the RNN model only used the previous count as a feature.
One approach could be to use the semantic information captured in texts.
\emph{Embeddings} are an established way to capture the semantics of texts (\cite{bengio_neural_nodate}, \cite{mikolov_efficient_2013}) and have been shown to improve the forecasting accuracy of a wide range of forecasting tasks~\cite{peng_transfer_2019}. 
Embeddings pre-trained on scientific texts~\cite{belt_scibert_2019} could be employed to render textual semantic information digestible to this work's model and improve forecasting accuracy.

Furthermore, future work could evaluate the presented approach on other datasets. 
arXiv is only one of multiple publication platforms. 
Augmenting the arXiv with data from other platforms might improve the inference of the model, as the model now has more training samples to learn patterns from.

Additionally, future work can change the taxonomy and thus get more fine-grained resolution on technologies. 
As shown in the literature review, the taxonomy strongly influences the performance of count-based forecasting approaches.
We assumed the 148 evaluated arXiv subcategories to be distinct areas of knowledge. 
Nevertheless, some subcategories contain multiple topics and are very exhaustive (e.g., cs.LG), while others are more specialized subcategories.
In extensive subcategories, averaging effects might take place that hide development patterns. 
A different, data-driven taxonomy might uncover interesting areas of knowledge and disaggregate development patterns, allowing the model to learn and use them to forecast.
One recently published dataset that provides a data-driven taxonomy is OpenAlex.
Its taxonomy encompasses $65.000$ so-called concepts organized in a directed-acyclic graph.
Concepts are abstract ideas that scholarly works discuss.
In OpenAlex\footnote{\url{https://docs.openalex.org/}}, every scholarly work is tagged with multiple of these concepts.
Future work could investigate if a more fine-grained taxonomy improves the forecasting accuracy of the RNN approach.

\newpage
\printbibliography

\end{document}